\documentclass{article}
\usepackage{graphicx}
\usepackage{amssymb}
\usepackage{amscd}
\usepackage{amsmath}
\usepackage[english]{babel}
\usepackage[T2A]{fontenc}
\usepackage{EuScript}
\usepackage{eufrak}
\usepackage{float}

\newcommand{\bee}{\begin{eqnarray}}
\newcommand{\eee}{\end{eqnarray}}

\def\beq{\begin{equation}}
\def\eeq{\end{equation}}
\def\ba{\beq\begin{array}{c}}
\def\ea{\end{array}\eeq}
\def\be{\ba}
\def\ee{\ea}
\def\nn{\nonumber}

\newcommand{\labcd}[2]{\hbox to\textwidth{#1\dotfill #2}}

\hoffset= - 1.10in         

\begin{document}

\hfill INR/TH -06-2007

\hfill ITEP/TH -15/07

\centerline{\Large{{\bf  Materializing Superghosts }}}
\bigskip
\centerline{Victor Alexandrov$^{a}$,\ \  Dmitry Krotov$^{b}$,\ \
Andrei Losev$^{c}$,\ \  Vyacheslav Lysov$^{d}$}

\begin{center}
$^a${\small{\em P.N. Lebedev Physical Institute Theoretical Physics
Division Russian Academy of Sciences, }}\\
$^b${\small{\em
Institute for Nuclear Research of the Russian Academy of Sciences, }}\\
$^{b,c,d}${\small{\em
Institute of Theoretical and Experimental Physics, }}\\
$^d${\small{\em  L.D. Landau Inst. for Theor. Phys. Russian Academy
of Sciences
,}}\\

$^{a,b}${\small{\em
Moscow State University, Department of Physics,}}\\

$^d${\small{\em Moscow Institute of Physics and Technology State
University ,}}

\end{center}
\bigskip
\vspace{6cm}

\centerline{ABSTRACT}
\bigskip

We construct the off-shell BV realization of $\mathcal{N}=1$, $d=10$
SYM with 7 auxillary fields. This becomes possible due to
materialized ghost phenomenon. Namely, supersymmetry ghosts are
coordinates on a manifold $B$ of 10-dimensional spinors with pure
spinors cut out. Auxillary fields are sections of a bundle over $B$,
and supersymmetry transformations are nonlinear in ghosts. By
integrating out axillary fields we obtain on-shell supersymmetric BV
action with terms quadratic in antifields. Exactly this on-shell BV
action was obtained in our previous paper after integration out of
auxiliary fields in the framework of Pure Spinor Superfield
Formalism.

\thispagestyle{empty}
\bigskip
\newpage
\section{What is materialization?} Recent investigations of what we
should call by symmetry of the system reveal that all empirical
notions may be expressed in terms of the very powerful
Batalin-Vilkovysky (BV) formalism. Moreover, we believe that this
formalism should be considered as a guiding line in extension of the
very notion of what we call symmetric system. In our previous paper
\cite{we} we explained how BV treats what one may call on-shell
symmetry (with control of terms proportional to the equations of
motions) on the example of supersymmetry. In the present paper we
will explain the new phenomena - the materialization of the
superghosts - again in BV formalism. In one sentence it means
replacing
 \beq
  \delta\phi^i \rightarrow \epsilon^{a} v_{a}^{i}(\phi)
\eeq
 by
 \beq \label{non-linear transformation}
  \delta\phi^i \rightarrow v^i( \epsilon, \phi),
\eeq
 where $v$ is a homogeneous function of degree 1 in
$\epsilon$, not necessarily linear. If we consider $\epsilon$ as a
parameter (as in the standard "Lie group" way of treating
supersymmetry) non-linear expressions in $\epsilon$ have no sense.
In BV philosophy $\epsilon$ should be treated as a ghost rather than
as infinitesimal parameter. This difference is essential - nonlinear
functions of degree 1 are different from linear ones only if
$\epsilon$ is {\bf even}, it means that the ghost point of view
turns out to be interesting for {\bf supersymmetry}. Perhaps, that
is why nonlinear effects what we are discussing here were missed
before.

In which sense the transformations (\ref{non-linear transformation})
close?  Traditionally, people would advocate \beq \label{traditional
commutator} \Big\{ v^i( \epsilon,
\phi)\frac{\partial}{\partial\phi^i} , v^i( \epsilon',
\phi)\frac{\partial}{\partial\phi^i} \Big\} =(\epsilon \gamma^\mu
\epsilon') P_\mu \eeq Here $P_\mu$ is the generator of translations,
and $\gamma_\mu$ are standard $\gamma$-matrices. Here we stress that
the proper expression is a little bit different

\beq \label{ghost materialization commutator} \Big\{ v^i( \epsilon,
\phi)\frac{\partial}{\partial\phi^i} , v^i( \epsilon,
\phi)\frac{\partial}{\partial\phi^i} \Big\} =(\epsilon \gamma^\mu
\epsilon) P_\mu
 \eeq
(there is no $\epsilon'$ in (\ref{ghost materialization
commutator})). Expression (\ref{ghost materialization commutator})
is the {\bf essence} of ghost materialization  approach.

Note, that (\ref{traditional commutator})  and (\ref{ghost
materialization commutator}) are equivalent for linear
transformations while they are not equivalent for nonlinear ones.
That is why the systems of equations coming from (\ref{traditional
commutator}) and (\ref{ghost materialization commutator}) are
different. Authors of \cite{Berk} find 9-dimensional space of
supersymmetries, following the standard approach (\ref{traditional
commutator}). While following the materialized ghost approach it is
possible to reconstruct the full 16-dimensional space of solutions.

Nonlinear equations arising  from (\ref{ghost materialization
commutator}) mean quite an interesting thing - the ghost $\epsilon$
turns out to be a point on a manifold of nontrivial topology rather
than a linear coordinate  on the algebra (with reversed parity).

That is why ghosts enter the game on the equal footing with the
matter fields, they are bosonic and span the manifold of possibly
nontrivial topology (i.e. they are getting shape). That is why we
are calling this phenomenon {\bf materialization}.

\subsection{BV approach to the notion of symmetry}
According to BV approach the symmetric system means that the BV
action $S(\Phi, \Phi^*,c,c^*)$ of the special form solves the master
equation
$$
 \frac{\delta S}{\delta c}\frac{\delta S}{\delta c^*}\ +\  \frac{\delta S}{\delta\Phi}\frac{\delta S}{\delta \Phi^*}\  =\ 0
$$
Note, that before indication of the exact form of $S$ the very
distinction of variables on ghosts and matter fields is meaningless.

  The standard {\bf off-shell symmetric system} means just that
$$
 S^{\hbox{\footnotesize{off}}}(\Phi, \Phi^*, c, c^*)\ =\ S^m(\Phi)\
+\  c^a v_{a}^{i}(\Phi)\Phi_{i}^*\  +\  f_{ab}^{c} c^a c^b c_{c}^*
$$
Here $S^m$ is the invariant action, $f_{a b}^c$ are structure
constants of the algebra and $v$ - vector fields, representing the
Lie (super)algebra.

The {\bf on-shell symmetric system} is just
$$
S^{\hbox{\footnotesize{on}}}(\phi, \phi^*, c, c^*)\ =\ S^m(\phi)\ +\
c^a v_{a}^{i}(\phi)\phi_{i}^*\  +\ f_{ab}^{c} c^a c^b c_{c}^*\ +\
\pi_{ab}^{ij}c^a c^b \phi_{i}^* \phi_{j}^*\
$$
i.e. consists of terms, quadratic both in ghosts and in antifields
(the fields $\Phi$ and $\phi$ are different).

  The on-shell symmetry corresponds to the closeness
of symmetry algebra up to terms proportional to the equations of
motion. The common way to get on-shell symmetry is to start with the
off-shell symmetrical systems (with fields $\Phi$) and integrate out
auxiliary fields ($\phi_{aux}$), so that we will stay with the
on-shell fields $\phi$. However, it may happen that either the
off-shell formulation is difficult to find or it involves too many
(infinitely many) auxiliary fields.

 We are trying to find the set of auxiliary fields and supersymmetry
transformations on them, such that there will be no terms quadratic
in antifields in the BV action. It happens that this can be done
adding 7 auxiliary scalar fields to the classical part of BV action
\cite{Berk}. After that the action becomes linear in antifields,
however the supersymmetry part is non-linear in superghosts.
\begin{equation}
S^{\hbox{\footnotesize{GM}}}(\Phi, \Phi^*, c, c^*)\ =\ S^m(\Phi)\ +\
v(c, \Phi)^i \Phi_{i}^*\  +\  f_{ab}^{c} c^a c^b c_{c}^*
\end{equation}
We call this action the theory with {\bf ghost materialization}.
Surely, integrating out these auxiliary fields we get back the
on-shell action.

 Below we will demonstrate all this phenomena in the case of
celebrated $\mathcal{N}=1$, $d=10$ SYM theory.

\subsection{Peculiarities of ghost materialization in $\mathcal{N}=1$, $d=10$
SYM: vector bundle of auxiliary fields}

It turns out that for $\mathcal{N}=1$, $d=10$ SYM the ghost
materialization works not for all but for almost all ghosts that are
even 16-dimensional left spinors of $SO(10)$. Namely, the domain $B$
where is works is obtained from $\mathbb{C}^{16}$ by excluding pure
spinors, i.e. those that satisfy
 \begin{equation}
B=\mathbb{C}^{16}-P,\;\;\;P=\{ \varepsilon; ( \varepsilon \gamma^{\mu} \varepsilon) = 0 \}
 \end{equation}
This is a first manifestation of materialization phenomenon - ghosts
form a nontrivial manifold $B$ rather than the linear space.

Moreover, auxiliary fields are sections of the 7-dimensional vector
bundle $A \rightarrow B$, taking values in the 10-dimensional fields
in adjoint representation of the gauge group. We will denote them as
$G_i(x)$, $i=1,\ldots , 7$. The bundle $A$ is equipped with the
scalar product $(\cdot,\cdot)_A$ which will be described in the next
section. Here we will just show the final result - the BV action for
$\mathcal{N}=1$, $d=10$ SYM coupled to auxiliary fields and
materialized ghosts\footnote{We would like to emphasize that the BV
form on the fields $G$ and $G^*$ is not unity. The contribution into
BV equation is given by
$$
\frac{\delta_L S}{\delta G^\alpha}\Big(
\frac{1}{2}(\varepsilon\gamma^\mu\varepsilon)\gamma_\mu^{\alpha\beta}\
-\ \varepsilon^\alpha\varepsilon^\beta\Big)\frac{\delta_R S}{\delta
G^\beta}
$$ The definitions of the scalar products $(\cdot,\cdot)_{10}$ and $(\cdot,\cdot)_A$
can be found in the next section (see also appendix).}:
\begin{equation} \label{Materialized superghosts BV action}
\begin{split}
S^{MSG}= \int d^{10}x\;
\hbox{Tr}\Big(-\frac{1}{2}F^2_{\mu\nu}+i\psi\gamma^\mu D_\mu \psi\
+\ (G,G)_A \ -\ (D_{\mu}c)A^\ast_\mu+ {g\{\psi,c\}}\psi^\ast -g ([G,
c], G^*)_A + g ccc^\ast + \\ +\ i (\varepsilon \gamma^\mu \psi)
A^\ast_\mu\ -\frac{1}{2}(\varepsilon
\gamma^{\mu\nu}\psi^\ast)F_{\mu\nu}\
     + (G,\ \psi^\ast)_{10}\ +\ i
(G^*,\ \gamma^\mu D_\mu \psi)_{10} +\\ +\ \eta^\mu
{[(\psi^\ast\partial_\mu\psi)+A_\nu^\ast
\partial_\mu A^\nu +c^\ast\partial_\mu c + (G^\ast , \partial_\mu G)_A ]}
+ i\eta_\mu^* (\varepsilon\gamma^\mu \varepsilon)\ +\ ic^\ast A_\mu
(\varepsilon \gamma^\mu \varepsilon) \Big)
\end{split}
\end{equation}
Integrating out auxiliary fields we get the on-shell action
\cite{we} with the terms quadratic in antifields
$$
-\frac{1}{8}(\varepsilon\gamma^\mu\varepsilon)(\psi^*\gamma_\mu\psi^*)
+ \frac{1}{4}(\varepsilon\psi^*)^2
$$
that was obtained in the approach of \cite{we} using the action
\begin{equation}\label{Fundamental theory with superghosts}
 S^{SUSY} = \int\ \hbox{Tr}\Big( <\EuScript{P},\
(Q+\Phi) \EuScript{A}> \ +\  g <\EuScript{P},\  \EuScript{A}^2\!>\ +
\ \sqrt{\!\!-i}\!\!<\EuScript{P},\ \varepsilon^\alpha  Q^s_\alpha
\EuScript{A}>\ + \ \!<\EuScript{P},\ \eta^\mu P^{s}_\mu
\EuScript{A}>\  + \ i\eta_\mu^*(\varepsilon\gamma^\mu\varepsilon)
\Big)
\end{equation}
with subsequent $Z_2$ projection (for details see \cite{we}). This
action is of the standard superfield type, however, it contains
infinitely many auxiliary fields and $Z_2$ symmetry of the effective
action is accidental for the present understanding. That is why we
use the materialized ghost approach in this paper. We do believe
that the geometry of the materialized ghost approach would show up
in the study of supergravities, where ghosts are promoted to fields.

\section{Geometry of the A-bundle}
Let us start with the trivial vector bundle $\mathbb{C}^{16} \rightarrow B$
and define its subbundle A as the space of solutions to equations:
\begin{equation} \label{bundle}
A=\{a\in \mathbb{C}^{16},\;\;(a \gamma^\mu\varepsilon)=0\}
\end{equation}
Here we have 10 equations on 16 variables, so naively, we may expect
6-dimensional space of solutions. However, one may show that for
$\varepsilon \in B $ the space of solutions is really 7 dimensional,
and this dimension jumps for pure spinors. Therefore, pure spinors
are excluded from the base.

For spinors from the base we will consider the 10-dimensional vector
$V(\varepsilon)$:
 $$
 V^{\mu}\ =\ (\varepsilon \gamma^\mu\varepsilon)
$$
From the Fiertz identity on 10-dimensional gamma-matrixes
\be%
\label{Fiertz first identity}%
(\gamma^{\mu})_{\alpha\beta}(\gamma_{\mu})_{\delta\sigma}\ =\
-\frac1 2 (\gamma^a)_{\alpha\delta}(\gamma_a)_{\beta\sigma} -
\frac1{24}(\gamma^{abc})_{\alpha\delta}
(\gamma_{abc})_{\beta\sigma}. \ee we obtain that $V$ is lightlike:
$$
V^{\mu} V^{\mu}\ =\ 0,
$$
also
\begin{equation} \label{gminus}
V^{\mu} \gamma^{\mu}\varepsilon \  =\ 0
\end{equation}
and for any element $a$ of the bundle A
\begin{equation} \label{aminus}
 V^{\mu} \gamma^{\mu} a\ =\ 0
\end{equation}
To see this one should contract the r.h.s. of (\ref{Fiertz first
identity}) with $\varepsilon^\alpha\varepsilon^\delta a^\beta$.
Define a trivial bundle $E$ as
$$
E\ =\ \{\ \varepsilon \in \mathbb{C}^{16}\ \}
$$
Therefore, both the bundle $A$ and  line bundle $E$ are subbundles
of the 8-dimensional bundle $C$ given by:
\begin{equation} \label{cbundle}
C\ =\ \{\ s\in \mathbb{C}^{16}\ ,\ \;\; (\varepsilon
\gamma^\mu\varepsilon) \gamma^{\mu}s  =0\}
\end{equation}
For the proof that the bundle $C$ is 8-dimensional see appendix.
Moreover, the bundle $C$ comes equipped with the non-degenerate
scalar product
$$
(s_1, s_2)_C=( s_1 , \gamma^{\mu} U^{\mu} s_2)_{10}
$$
where for any non-pure spinor $\varepsilon$ we define vector
$U^\mu$, such that \be U^\mu V^\mu=1 \ee and by $(\cdot,\cdot)_{10}$
we denote the standard bilinear paring on 16-component $SO(10)$
spinors (for the definition in terms of anticommuting variables see
appendix) \be (\cdot,\cdot)_{10}:\;\; S^R\otimes S^L \rightarrow
\mathbb{C} \ee From the construction  it is clear that $A$ is just
the orthogonal complement to $E$ in $C$: \be \nn
C=E\oplus A\\
a \in A, \varepsilon \in E: (a,\varepsilon)_C=0 \ee Therefore, the
pairing $(\cdot,\cdot)_C$ induces on $A$ the pairing
$(\cdot,\cdot)_A$.

\section{Relation to Berkovits proposal}
In order to make contact with Berkovits proposal started in \cite{Berk}
 we will consider a patch in the base $B$ where the bundle $A$
may be trivialized. We will look for the orthogonal basis in $A$
that we will call  $\upsilon_i(\varepsilon)$ considered as elements
of $\mathbb{C}^{16}$, and write
$$
G\ =\ G_i(x)\upsilon_i(\varepsilon), \ \ \ \ \ \ \ \ \ G^*\ =\
G_i^*(x)\upsilon_i(\varepsilon)
$$
therefore, the action will take the form
\begin{equation} \label{SO(7) effective action}
\begin{split}
S^{MSG}\ =\ \int d^{10}x\;
\hbox{Tr}\Big(-\frac{1}{2}F^2_{\mu\nu}+i\psi\gamma^\mu D_\mu \psi\
+\ G_i^2\ -\ D^{\mu}cA^\ast_\mu + {g\{\psi,c\}}\psi^\ast\ - g[G^i,
c]G_i^* + g ccc^\ast\ + \\ +\ i (\varepsilon \gamma^\mu \psi)
A^\ast_\mu
-\frac{1}{2}(\varepsilon\gamma^{\mu\nu}\psi^\ast)F_{\mu\nu}\
     + G_i\upsilon_i \psi^\ast -i
\upsilon_i\gamma^\mu D_\mu \psi G_i^\ast +\eta^\mu
{[(\psi^\ast\partial_\mu\psi)+A_\nu^\ast
\partial_\mu A^\nu +c^\ast\partial_\mu c + G_i^\ast\partial_\mu G_i]}
\ +\\+\  i\eta_\mu^* (\varepsilon\gamma^\mu\varepsilon)\ +\ ic^\ast
A_\mu (\varepsilon\gamma^\mu\varepsilon) \Big)
\end{split}
\end{equation}
The classical invariant action is given by the first three terms in
the first line. The supersymmetry transformations can be extracted
from the first fourth terms of the second line. They are exactly
those  suggested in \cite{Berk}. The last term in the action
reflects \cite{we} the fact that we a working in a certain gauge
(analog of Wess-Zumino gauge), thus the commutator of two
supersymmetry transformations is closed only up to a gauge
transformation with parameter
$(\varepsilon\gamma^\mu\varepsilon)A_\mu$. This action (\ref{SO(7)
effective action}) can be considered as an off-shell BV formulation
of $\mathcal{N}=1$, $d=10$ SYM.

From the definition of $\upsilon$ it follows  that
 \be\label{eq}
\upsilon^i\gamma^\mu\varepsilon=0\\
\upsilon^i\gamma^\mu\upsilon^k-\delta^{i
k}(\varepsilon\gamma^\mu\varepsilon)=0 \ee Note, that from
completeness of the basis it follows that
 \be\label{completeness}
 \sum\limits_i
\upsilon^\alpha_i\upsilon^\beta_i=\frac{1}{2}(\varepsilon\gamma^\mu\varepsilon)
\gamma_\mu^{\alpha\beta}-\varepsilon^\alpha\varepsilon^\beta \ee

It should be mentioned that the  natural group of symmetry of the
system (\ref{eq}) and completeness identity (\ref{completeness}) is
not $SO(7)$ but $SO(8)$. Namely it is natural to unite ghosts
$\varepsilon$ and $\upsilon^i$ into a single $SO(8)$ multiplet $u^A\
=\ (\ \varepsilon, \ \upsilon^i)$. The system (\ref{eq}) can be
written then as
\begin{equation}\label{SO(8) system of equations}
u^A\gamma^\mu u^B\ =\ \frac{1}{8} \delta^{A B} (u^C\gamma^\mu u_C)
\end{equation}
In supergravity $\varepsilon$ becomes a field. Thus, probably there
will be a local symmetry mixing the ghosts $\varepsilon$ and
$\upsilon^i$.

Integrating out auxiliary fields $G_i$ from the action (\ref{SO(7)
effective action}) on the lagrangian submanifold $G_i^*\ =\ 0$, one
can obtain the on-shell BV action
\begin{equation}
\begin{split} \label{on-shell BV action} S^{\hbox{\footnotesize{on-shell}}}= \int d^{10}x\;
\hbox{Tr}\left(-\frac{1}{2}F^2_{\mu\nu}+i\psi\gamma^\mu D_\mu \psi
-D^{\mu}cA^\ast_\mu+ g {\{\psi,c\}}\psi^\ast+ g ccc^\ast+ i
(\varepsilon \gamma^\mu \psi) A^\ast_\mu
-\frac{1}{2}(\varepsilon\gamma^{\mu\nu}\psi^\ast)F_{\mu\nu}\right.\\
\left. + \eta^\mu {[(\psi^\ast\partial_\mu\psi)+A_\nu^\ast
\partial_\mu A^\nu +c^\ast\partial_\mu c]} +i\eta_\mu^*
(\varepsilon\gamma^\mu\varepsilon)+ic^\ast A_\mu
(\varepsilon\gamma_\mu\varepsilon)
-\frac{1}{8}(\varepsilon\gamma^\mu\varepsilon)(\psi^*\gamma_\mu\psi^*)
+ \frac{1}{4}(\varepsilon\psi^*)^2\right)
\end{split}
\end{equation}
found in \cite{we} integrating out auxiliary fields from the
superfield-like  action (\ref{Fundamental theory with superghosts}).

Now we are coming to the main point - what is the difference between
our approach and approach of \cite{Berk}? The difference is in the
setup. In \cite{Berk} it was found linear solution to the system
(\ref{eq}). Such solution can not be found for the full
16-dimensional space. This is clear from the structure of relations
on pure spinor constraints. Namely, in the paper \cite{we-first} it
was shown that the $Q$-cohomologies can be calculated using the
tower of fundamental relations. In case of 10-dimensional quadrics
the unique system of these relations is given by
$$
\begin{array}{c}
f^\mu\ =\ \varepsilon\gamma^\mu\varepsilon\\
G^\mu_\alpha\ =\ (\varepsilon\gamma^\mu)_\alpha\\
G^{\alpha\beta}\ =\ \varepsilon^\alpha\varepsilon^\beta\ -\
\frac{1}{2}(\varepsilon\gamma^\mu\varepsilon)(\gamma^\mu)^{\alpha\beta}\\
G^\mu_\alpha\ =\ (\varepsilon\gamma^\mu)_\alpha\\
f^\mu\ =\ \varepsilon\gamma^\mu\varepsilon
\end{array}
$$
This means that the following relations take place: $G_\alpha^\mu\
f_\mu\ =\ 0$,\ \ \ $G^{\alpha\beta}\ G^\mu_\beta\ =\ 0$, etc. These
relations are valid without imposing pure spinor constraints $f^\mu\
=\ 0$. Appearance of 5 relations leads to 6 well known
representatives of cohomologies.

Be there a linear solution to the system (\ref{eq}), the first
equation
$$
\upsilon^i(\varepsilon)\gamma^\mu\varepsilon=0
$$
states that there is another relation on $G_\alpha^\mu$, which is
linear in $\varepsilon^\alpha$, hence does not coincide with
$G^{\alpha\beta}$. If that is true, the cohomologies of SYM would be
different. Thus, there is no such linear solution.

The situation is different in $d=4,6$, where we do have a linear
dependence $\upsilon^i(\varepsilon)$ but the structure of
cohomologies for SYM is also different \cite{hol}.

Another argument is the non-triviality of the $A$ bundle
\cite{Golovko}. Be there  linear solution to the system (\ref{eq}),
this would imply that the  bundle $A$ is trivial. However, we expect
that the non-triviality of this bundle can be measured by the
corresponding Pontryagin class. All this is in accord with
\cite{Berk}, were the linear solution was found only on the
9-dimensional subspace of the full 16-dimensional one, where the
solution can be linearized.

\section{Appendix}
The algebra of $\gamma$-matrices
$$
\{\gamma^\mu,\ \gamma^\nu\}\ =\ 2g^{\mu\nu}
$$
can be represented by differential operators
$$
\gamma^\mu\ =\ (\xi^\mu\ +\ \frac{\partial}{\partial\xi^\mu})\ , \ \
\ \ \ \ \ \ \ \mu\ =\ 1\ .\ .\ .\ 5
$$
\vspace{-0.5cm}
$$
\gamma^{\mu+5}\ =\ (-i)(\xi^\mu\ -\
\frac{\partial}{\partial\xi^\mu})\ , \ \ \ \ \{\xi^\mu,\xi^\nu\}\ =\
0\ \ \
$$
of odd variables $\xi^\mu$. These operators act between the spaces
of left and right spinors which are odd and even functions of
$\xi^\mu$ respectively. This space is equipped with the Lorentz
invariant scalar product
$$
(\psi,\ \chi)_{10}\ =\ \int d\xi^1\ ...\ d\xi^5 i^{P(\chi)} \ \ \psi
\chi
$$
were $P(\chi)$ is the number of $\xi^\mu$ variables in the $\chi$.

To solve the system of equations
\begin{equation}\label{appendix system of equations}
u^A\gamma^\mu u^B\ =\  \delta^{A B} (\varepsilon\gamma^\mu
\varepsilon)
\end{equation}
which is equivalent to (\ref{eq}) and (\ref{SO(8) system of
equations}) one can use the fact that the vector
$(\varepsilon\gamma^\mu\varepsilon)$ is lightlike and choose the
frame were
\begin{equation}\label{light frame}
(\varepsilon\gamma^1\varepsilon)\ = \ 1,\ \ \ \ \ \
(\varepsilon\gamma^6\varepsilon)\ = \ i, \ \ \ \ \ \
(\varepsilon\gamma^I\varepsilon)\ = \ 0, \ \ \ \ \ I=2...5,7...10
\end{equation}
Using the definitions
$$
\gamma^+\ =\ \frac{1}{2}(\gamma^1\ +\ i\gamma^6),\ \ \ \ \ \
\gamma^-\ =\ \frac{1}{2}(\gamma^1\ -\ i\gamma^6)
$$
one can write the system (\ref{appendix system of equations}) as
$$
u^A\gamma_-u^B\ =\ \delta^{A B}
$$
$$
u^A\gamma_+u^B\ =\ 0
$$
$$
u^A\gamma_Iu^B\ =\ 0
$$
Since $\gamma^-\ =\ \frac{\partial}{\partial\xi^1}$, one can write
explicit solution of this system as
\begin{equation}\label{solution in terms of anticommuting variables}
u_A\ =\ \xi^1 P_A(\hat{\xi}), \ \ \ \ \ \hat{\xi}\ =\ \xi^2...\xi^5
\end{equation}
The space of $P_A$ is 8 dimensional. The corresponding orthonormal
basis elements are given by \be P^{+}=\frac{1}{\sqrt{2}}(1+
\xi^2\xi^3\xi^4\xi^5), \;\;\;P^{-}=\frac{i}{\sqrt{2}}(1-
\xi^2\xi^3\xi^4\xi^5)\nonumber\\
\nonumber \\ P_K^{+}=\frac{i}{\sqrt{2}}(\xi^2\xi^K+
\partial_K(\xi^3\xi^4\xi^5)), \ \ \
P_K^{-}=\frac{1}{\sqrt{2}}(\xi^2\xi^K-
\partial_K(\xi^3\xi^4\xi^5)), \ \ \ \ K=3,4,5 \nonumber \ee

The last thing to mention is that the $C$-bundle (\ref{cbundle}) is
8 dimensional. This is clear from its definition in the light-cone
frame (\ref{light frame})
$$
\gamma^+ s\ =\ 0
$$
The general solution of this equation is given by (\ref{solution in
terms of anticommuting variables}), hence is 8-dimensional.

\section{Acknowledgments}
 We would like to thank Alexei Gorodentsev for helpful comments. It is
a pleasure to thank S.~Demidov, A.~Rosly and V.~Rubakov for useful
critical discussions.  The work of DK was supported by the grant
RFBR-05-02-17363 and the fellowship of Dynasty Foundation in 2007.
The work of AL was supported by the grant RFBR 07-02-01161,  INTAS
03-51-6346, NWO-RFBR-047.011.2004.026 (RFBR 05-02-89000-NWOa) and
the grant for support of scientific schools NSh-8065.2006.2. The
work of VL was supported by the grant RFBR 07-02-01161 and INTAS
03-51-6346.

\end{document}